\documentclass[
reprint,
superscriptaddress,
nofootinbib,
amsmath,amssymb,
aps,
]{revtex4-1}

\usepackage{graphicx}
\usepackage{dcolumn}
\usepackage{bm}
\usepackage[colorlinks=true,allcolors=blue]{hyperref}
\usepackage[colorinlistoftodos]{todonotes}

\usepackage[normalem]{ulem} 
\newtheorem{theorem}{Theorem}[section]
\newtheorem{corollary}[theorem]{Corollary}
\newtheorem{lemma}[theorem]{Lemma}
\newtheorem{proposition}[theorem]{Proposition}

\usepackage{pgf,tikz}
\usepackage{multirow,booktabs}
\usepackage{textcomp} 
\usetikzlibrary{matrix, arrows, automata}
\usepackage{blkarray} 
\usepackage{esint} 
\usepackage{blindtext}
\usepackage{color}
\usepackage[normalem]{ulem} 

\usepackage{chngcntr} 

\newcommand{\chg}[2][red]{{\color{#1}#2}}

\newcommand\rem[1]{}

\renewcommand{\chg}[2][black]{#2}

\begin{document}

\title{The role of topology and mechanics in uniaxially growing cell networks}

\author{Alexander Erlich}
\email[To whom correspondence should be addressed. E-mail: ]{alexander.erlich@univ-grenoble-alpes.fr}
\affiliation{Laboratoire Interdisciplinaire de Physique (LIPhy), Universit\'e Grenoble Alpes, CNRS, Grenoble 38000, France}
\author{Gareth W. Jones}
\affiliation{School of Mathematics, University of Manchester, Oxford Road, Manchester M13 9PL, United Kingdom}
\author{Fran\c coise Tisseur}
\affiliation{School of Mathematics, University of Manchester, Oxford Road, Manchester M13 9PL, United Kingdom}
\author{Derek E. Moulton}
\affiliation{Mathematical Institute, University of Oxford, Andrew Wiles Building, Woodstock Road, Oxford OX2 6GG, United Kingdom}
\author{Alain Goriely}
\affiliation{Mathematical Institute, University of Oxford, Andrew Wiles Building, Woodstock Road, Oxford OX2 6GG, United Kingdom}

\begin{abstract}
	In \chg{biological} systems, the growth of cells, tissues, and organs is influenced by mechanical cues. Locally, cell growth leads to a mechanically heterogeneous en\-viron\-ment as cells pull and push their neighbors in a cell network. Despite this local heterogeneity, at the tissue level, the cell network is remarkably robust, as it is not easily perturbed by changes in the mechanical environment or the network connectivity. Through a network model, we relate global tissue structure (i.e. the cell network topology) and local growth mechanisms (growth laws) to the overall tissue response. Within this framework, we investigate the two main mechanical growth laws that have been proposed:  stress-driven or strain-driven growth. We show that in order to create a robust and stable tissue environment, networks with predominantly series connections are naturally driven by stress-driven growth, whereas networks with predominantly parallel connections are associated with strain-driven growth. 
\end{abstract}


\maketitle
\onecolumngrid

\section{Introduction}

\begin{figure*}
	\hfill{}\includegraphics[width=1\textwidth]{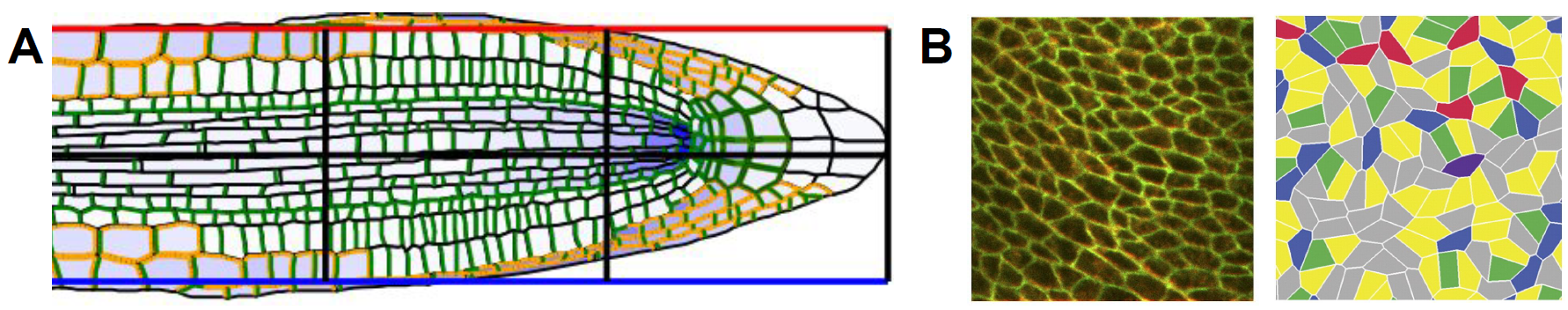}\hfill{}
	\caption{Vertex based models of biological tissues with different topologies. The cross-section the primary root of Arabidopsis thaliana (A) exhibits an almost brick-like structure of cells, growing primarily in the axial direction. The image shows a segmented image of the root, see \cite{fozard2016hybrid}. The topology of a cross-section of the apical layer of epithelial tissue is polygonal (B), with hexagons occurring most often, see \cite{farhadifar2007influence}.
	}
	\label{fig:vertex-models}
\end{figure*}

Many biological tissues take cues from their mechanical environment to regulate growth and, in turn, generate mechanical stresses on their surrounding \cite{hu01}. At the tissue level, arteries respond to wall shear stress \cite{taber2001biomechanics} and  skeletal muscle is primarily driven by mechanical stretch \cite{zollner2012stretching}, which has been successfully described by the theory of morphoelasticity that combines growth and  remodeling with large mechanical deformations \cite{goriely17,kuhl2014growingMatter,amatar10,goriely2019growth}. 
However, many biological tissues exhibit a cellular structure, and vastly different network topologies are observed in nature, e.g. cells in epithelial monolayers are polygonal with hexagons occurring most frequently in planar slices (Fig. \ref{fig:vertex-models}A, \cite{farhadifar2007influence}), whereas plant root cells form networks of cuboids like bricks in a wall (Fig. \ref{fig:vertex-models}B, \cite{band2012growth}). Bespoke mechanical modelling is required to capture such inherently discrete geometric structures as they grow. 

From a mechanical point of view, cell growth causes a local volumetric and mass increase, pushing and pulling neighboring cells and thus leading to a local buildup of mechanical stress. Such geometric and mechanical confinement of a cell in turn adapts and regulates cell growth. The mutual feedback mechanism between cell growth and mechanical and other fields is called a \emph{growth law}, and the tissue evolution is termed \emph{growth dynamics}. Mechanical growth laws have been studied both for patterning processes in developing tissues \cite{odell1981mechanical,shraiman2005mechanical} and regulatory processes in mature tissues \cite{taber2001biomechanics}, both with discrete and continuum models. Their forms have been either phenomenologically and micro-structurally inspired \cite{goktepe2010athlete,taber2008theoretical} or based on thermodynamical arguments \cite{ambrosi2007stress}. The main classification is whether growth is \textit{stress-driven} or \textit{strain-driven}  \cite{goktepe2010multiscale,goriely17}. 

When connecting strain-driven or stress-driven growing cells into a network structure, growth dynamics at the network level can become extremely complex \cite{shraiman2005mechanical,taber2008theoretical}. Growth dynamics is highly dependent on the form of the growth law \cite{vago09}. For a given network topology and growth law, it is unclear if growth dynamics will drive the cell network towards an equilibrium state and provide homeostasis \cite{cyron14,latorre2018mechanobiological},  grow without bounds, or shrink to oblivion \cite{erlich18}. 

Here, we investigate the role of both the growth law and network topology in homeostasis, in particular how the topology of the network impacts the stability of an equilibrium state if the underlying growth law is stress-driven or strain-driven. \chg{While most biological cell networks are two- or three-dimensional, we study here, as a first step, a network deforming along a single axis. The advantage of this approach is that the analytical tractability enables concrete connections to be made between topology, the form of growth law, and stability and thus leads to a broad understanding of the relative role of mechanics and topology in the dynamics of growth, albeit in a simplified geometry.} We note that similar questions have been explored in the context of solute transport in networks, e.g. how the vessels of a network adapt to optimize transport \cite{ronellenfitsch2016global}, or how network topology dynamically adapts to improve transport \cite{marbach2016pruning}. Our approach is  different from the homogenization of micromechanical network models, which are typically limited to (near-) periodic lattices \cite{chenchiah2014energy,murisic2015discrete}. Instead, we combine the dynamics of growth with  the algebraic structure of an inherently discrete and, possibly, disordered network. 

\section{\label{sec:growth-dynamics-of-a-single-cell}Growth dynamics of a single cell}

To introduce the concept of growth dynamics, consider first a single cell in isolation, which deforms only along one axis. Its deformed length is $l$, and its rest length is $L$. The force $T=h(\alpha)$ in the cell along the growth axis is related to the elastic stretch $\alpha = l/L$ by a monotonic non-linear constitutive law $h$ and assumed to be homogeneous. A simple strain-driven growth law, $\dot{L}=K_{\omega}(l-l^{*})$, dictates that the cell grows when its length is different from the equilibrium length $l^{*}$  with a growth constant $K_\omega$. Assuming that a constant  external force is prescribed on the cell, $T=F_\mathrm{ext}$. Hence we have
\begin{equation}
\dot{L}=K_{\omega}(\alpha_{\text{ext}}L-l^{*}),
\end{equation} 
where $\alpha_{\text{ext}}=h^{-1}(F_\mathrm{ext})$ and  the growth dynamics will converge to the equilibrium state $L^{*}=l^{*}/\alpha_{\text{ext}}$ if and only if the growth constant $K_\omega$ is negative. In that case, a cell with length shorter than the equilibrium length, $l<l^{*}$,  grows ($\dot{L}>0$) until the equilibrium length is reached. In the case of stress-driven growth, a linear growth dynamic reads 
\begin{equation}
\dot{L}=K_{\tau}(T-T^{*}).
\end{equation}
Similar reasoning shows that under an externally prescribed cell length, the growth constant $K_\tau$ must be positive for the growth dynamics to converge to an equilibrium state. 


\section{\label{sec:growth-dynamics-of-a-cell-network}Growth dynamics of a network of cells}

\begin{figure*}
	\centering
	\includegraphics[width=1\textwidth]{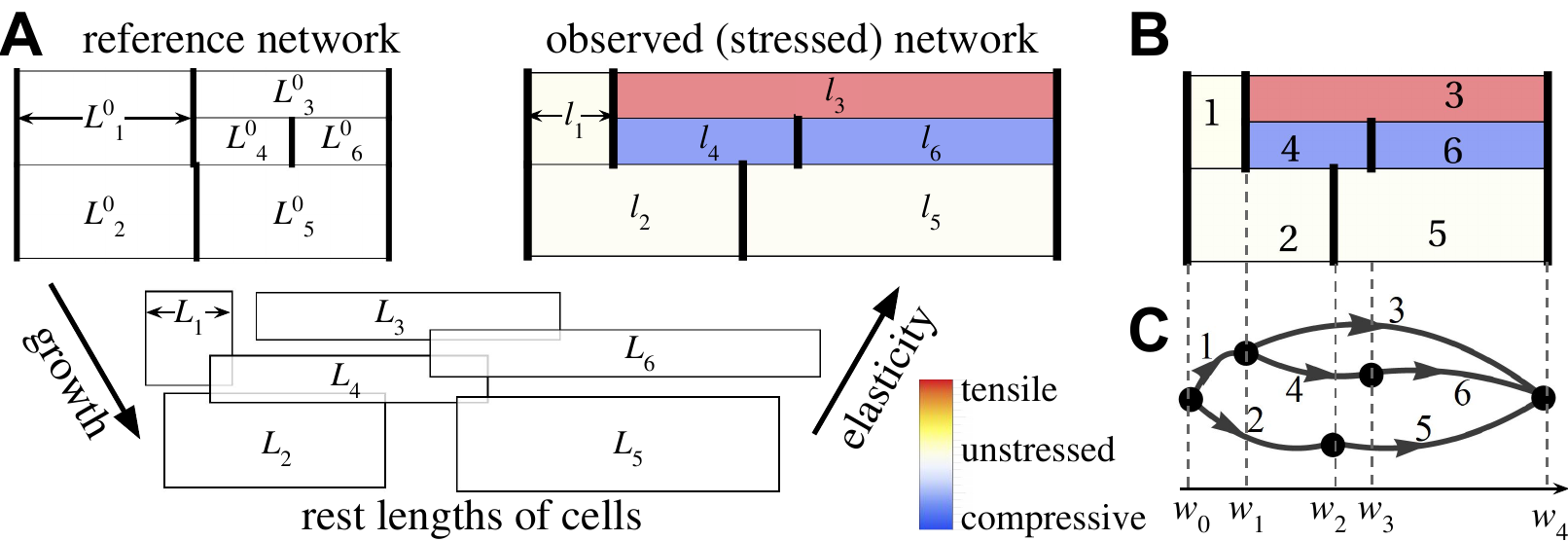}
	\caption{\label{fig:mechanical-and-geometric-setup} Model setup. \textbf{A} 
		When individual cells grow at different rates, the overall network size increases but creates internal stresses  due to geometric incompatibility. For example, if cell 3 grows by a factor $L_3/L_3^0>1$, it is subjected to an elastic stretch $l_3/L_3>1$ due to pulling from its neighbors and the cell stress  is  tensile,  $T_3=h_3(\alpha_3)>0$. \textbf{B} The topology of the cell network can be encoded by a  graph, where nodes correspond to vertical (force-bearing) walls and edges correspond to cells. 
		\label{fig1:morphoelastic-bar}} 
\end{figure*}

While the stability of growth dynamics for an isolated cell is determined by the sign of the growth constant, the situation is more complicated when multiple cells are organized into networks, which impose additional constraints through length and force balances. The question is then to establish if the entire network will exhibit homeostasis, i.e. whether small perturbations around an equilibrium state will lead the tissue back to its  equilibrium. We investigate here the relationship between network topology and growth laws in a highly idealized geometry of cuboid cells connected at their walls and only allowed to deform along one axis. We also note the analogy between a network of cells and  electrical circuits;  laws such as Kirchhoff's circuit  laws have a natural equivalent in the mechanical network.

We consider a network of $m$ cells which deform along the horizontal $x$-axis as shown in Fig.~\ref{fig1:morphoelastic-bar}B. The network  topology is described by a directed graph in which $m$ edges represent cells and $n$ nodes represent vertical, force-bearing walls. The leftmost wall is not  (not counting the leftmost wall) \cite{strang1993introduction,grady2010discrete}. Let $\mathsf{B}\in\{-1,0,1\}^{m\times n}$ be the network's oriented incidence matrix ($B_{ij}=+1$ if the edge $j$ starts at vertex $i$, $-1$ if it arrives at vertex $i$, 0 if it does not connect vertex $i$). Not including the leftmost wall in $\mathsf{B}$ ensures that the network is fixed in space, making the graph Laplacian $\mathsf{B}^T \mathsf{B}$ regular. The force balance at each vertical wall (shown in bold in Fig.~\ref{fig:mechanical-and-geometric-setup}) is given by
\begin{equation}
\mathsf{B}^{T}\boldsymbol{T}=\boldsymbol{T}_{\mathrm{ext}}
\label{eq1:KCL}
\end{equation}
where $\boldsymbol{T}=(T_{1},\ldots,T_{m})^{T}$ describes the forces of cells along the $x$-axis and $\boldsymbol{T}_{\mathrm{ext}}$ describes the externally prescribed forces at the walls. Eq.~\eqref{eq1:KCL} is equivalent to Kirchhoff's current law if the currents are replaced by the forces. The current lengths of cells are  $\boldsymbol{l}=(l_{1},\ldots,l_{m})^{T}$ and their unstressed lengths are $\boldsymbol{L}=(L_{1},\ldots,L_{m})^{T}$. Not counting the leftmost wall, cells are separated by $n$ walls with horizontal coordinates $\boldsymbol{w}=(w_{1},\ldots,w_{n})^{T}$, where $w_{n}$ is the rightmost wall and we have $\boldsymbol{l}=\mathsf{B}\boldsymbol{w}$. We assume a constitutive relationship between the elastic stretch $\alpha_i=l_i/L_i$ of a cell and its force response $h_i$:
\begin{equation}
T_{i}=h_{i}(\alpha_{i}),\qquad h_{i}'(\alpha_{i})>0,\qquad h_{i}(1)=0.
\label{eq2:constitutive-single-cell}
\end{equation}
Typical laws are neo-Hookean model, $h(\alpha)=\mu (\alpha^2-\alpha^{-1})$, or the Hencky model $h(\alpha) =\mu  \log \alpha$ \cite{mihai2017characterize}. 
Growth dynamics is modeled by the evolution of the rest lengths $L_i(t)$. We define the stress-driven growth law 
\begin{equation}
\dot{L}_{i}  =\mathcal{G}_{\tau}(T_{i})
\end{equation}
to be a function of only the force $T_{i}$, and the strain-driven growth law takes the form 
\begin{equation}
\dot{L}_{i} =\mathcal{G}_{\omega}(l_{i}),
\end{equation}
see \cite{goriely17}.
The governing equations of growth dynamics combine the force balance \eqref{eq1:KCL}, the constitutive relationships \eqref{eq2:constitutive-single-cell} and either the stress-driven or strain-driven growth laws.

Boundary conditions are either fixed total length or imposed force for the entire network. \chg{We note that the fixed length condition is the discrete equivalent to a Dirichlet (i.e. prescribed displacement) condition in a continuous system. Likewise prescribed force is equivalent to a Neumann (i.e. prescribed traction) condition.} If a force $F_\mathrm{ext}$ is prescribed on the outermost walls, then $\boldsymbol{T}_\mathrm{ext}=(0,\ldots,0,F_\mathrm{ext})^T$. If instead the total length $L_\mathrm{ext}$ is prescribed, the dimension of the incidence matrix is reduced to $m\times(n-1)$, as the position of the rightmost coordinate $w_n=L_\mathrm{ext}$ is not an unknown. In this case $\boldsymbol{T}_\mathrm{ext}$ is a zero vector. In the case of linear stability analysis, both boundary conditions can be taken into account by only changing the dimensions of the incidence matrix. Thus, without loss of generality \chg{for stability analysis}, we focus on the force prescribed case.

\begin{figure}
	\centering
	\includegraphics{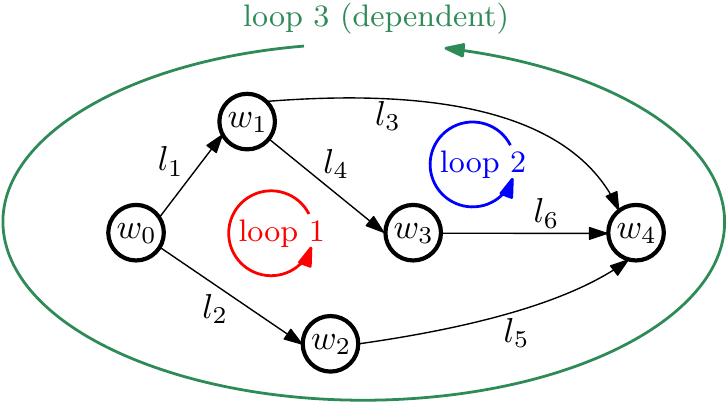}
	\caption{Illustration of the loops of an example network, and their relationship to the nullspace of $\mathsf{B}^{T}$. Two of the three loops shown here are independent, which is reflected by the fact that the nullspace of $\mathsf{B}^{T}$ is two dimensional. The choice of basis vectors in (\ref{eq:nullspace-and-loops}) corresponds to loops 1 and 2, and the fact that loop 3 is a dependent loop is reflected by the fact that it can be described by a linear combination of the basis vectors of loops 1 and 2. }
	\label{fig:loops}
\end{figure}

\section{\label{sec:worked-example}Length and force balance of an example network}
To better illustrate how forces and lengths are balanced in our framework, and to develop a better intuition for how cell graphs and cell diagrams relate to each other, we state the incidence matrix and resulting balances for the example network shown in Figure \ref{fig1:morphoelastic-bar}C. In this example, there are $n=4$ vertical walls (not counting the leftmost wall) and $m=6$ cells. The transpose of the incidence matrix is 
\begin{equation}
\mathsf{B}^{T}=\left(\begin{array}{cccccc}
\phantom{-}1 & \phantom{-}0 & -1 & -1 & \phantom{-}0 & \phantom{-}0\\
\phantom{-}0 & \phantom{-}1 & \phantom{-}0 & \phantom{-}0 & -1 & \phantom{-}0\\
\phantom{-}0 & \phantom{-}0 & \phantom{-}0 & \phantom{-}1 & \phantom{-}0 & -1\\
\phantom{-}0 & \phantom{-}0 & \phantom{-}1 & \phantom{-}0 & \phantom{-}1 & \phantom{-}1
\end{array}\right)\,.
\label{eq:example-incidence-matrix}
\end{equation}
The force balance at each vertical wall is given by \eqref{eq1:KCL}. With no externally prescribed forces ($\boldsymbol{T}_\mathrm{ext}=\boldsymbol{0}$), the force balances at walls 1--4 are
\begin{equation}
\mathsf{B}^{T}\boldsymbol{T}=\begin{pmatrix}T_{1}-T_{3}-T_{4}\\
T_{2}-T_{5}\\
T_{4}-T_{6}\\
T_{3}+T_{5}+T_{6}
\end{pmatrix}=\begin{pmatrix}0\\
0\\
0\\
0
\end{pmatrix}\quad\begin{array}{c}
\text{at }w_{1}\\
\text{at }w_{2}\\
\text{at }w_{3}\\
\text{at }w_{4}\,.
\end{array}
\end{equation}

In our example, lengths $\boldsymbol{l}$ are related to wall coordinates by $\boldsymbol{l}=\tilde{A}\boldsymbol{w}$, which has components

\begin{equation}
l_{1}=w_{1},\quad l_{2}=w_{2},\quad l_{3}=w_{4}-w_{1},\quad l_{4}=w_{3}-w_{1},\quad l_{5}=w_{4}-w_{2},\quad l_{6}=w_{4}-w_{3}.
\end{equation}

A basis for the null space of $\mathsf{B}^T$ is given by the vectors 
\begin{equation}
\boldsymbol{\xi}_{1}=(-1, 1, 0, -1, 1, -1)^{T},\qquad\boldsymbol{\xi}_{2}=(0, 0, -1, 1, 0, 1)^{T}.
\end{equation}
As noted before, the components $+1$ and $-1$ indicate whether the direction goes with or against the arrow, hence $\boldsymbol{\xi}_{1}$ corresponds to the loop 1 (red in Figure \ref{fig:loops}) which comprises segments $l_1$, $l_2$, $l_4$, $l_5$ and $l_6$. Similarly, $\boldsymbol{\xi}_{2}$ corresponds to loop 2 (blue in Figure \ref{fig:loops}) which comprises segments $l_3$, $l_4$ and $l_5$. The number of independent loops is $\dim N(B^{T})=m-n+1=2$. Note that the outer big loop 3 (green in Figure \ref{fig:loops}) which comprises segments $l_1$, $l_2$, $l_3$ and $l_5$ is represented by the linear combination $\boldsymbol{\xi}_{1}+\boldsymbol{\xi}_{2}=(-1, 1, -1, 0, 1, 0)^{T}$ and is therefore not an independent loop. 
In summary, the length balance according to Kirchhoff's voltage law is 
\begin{equation}
\underbrace{\boldsymbol{\xi}_{1}\cdot\boldsymbol{l}=-l_{1}+l_{2}-l_{4}+l_{5}-l_{6}=0}_{\text{loop 1}},\qquad\underbrace{\boldsymbol{\xi}_{2}\cdot\boldsymbol{l}=-l_{3}+l_{4}+l_{6}=0}_{\text{loop 2}}.
\label{eq:nullspace-and-loops}
\end{equation}

\section{Mechanical stability of cell assemblies}

The goal is to identify the relevant parameters for which the growth laws are stable, and to quantify the probability of their stability for these parameters given a randomly assembled network. We assume that an equilibrium state $\boldsymbol{w}^*$, $\boldsymbol{T}^*$, $\boldsymbol{L}^*$ of the governing equations exists, satisfying $(\mathcal{G}_{\omega})_{i}(l_{i}^{*})=0$ in the strain-driven case and $(\mathcal{G}_{\omega})_{i}(l_{i}^{*})=0$ in the stress-driven case. We keep the form of stress-strain relationships $h_i$ and the functional form of the growth law $\mathcal{G}_i$ general. With a small parameter $\varepsilon$, we expand

We start with a homeostatic equilibrium state  ($\boldsymbol{w}^*$, $\boldsymbol{T}^*$, $\boldsymbol{L}^*$ for which $\dot{L}_i=0$ for all  $i$) and  perform a linear stability analysis by expanding

\begin{equation}
\boldsymbol{w}\approx\boldsymbol{w}^{*}+\varepsilon\boldsymbol{\omega},\quad\boldsymbol{T}\approx\boldsymbol{T}^{*}+\varepsilon\boldsymbol{\tau},\quad\boldsymbol{L}\approx\boldsymbol{L}^{*}+\varepsilon\boldsymbol{\Lambda}.
\label{eq:linear-stability-ansatz}
\end{equation} 

Using the series expansion \eqref{eq:linear-stability-ansatz} in the governing equations \eqref{eq1:KCL}, \eqref{eq2:constitutive-single-cell} and either the stress-driven growth law $\dot{L}_{i}  =\mathcal{G}_{\tau}(T_{i})$ or the strain-driven growth law $\dot{L}_{i} =\mathcal{G}_{\omega}(l_{i})$, leads to the following system at linear order in $\varepsilon$:
\begin{equation}
\underbrace{\begin{pmatrix}\mathsf{B}^{T} & 0 & 0\\
	\mathsf{I}_{m} & -\mathsf{H}_{\omega}\mathsf{B} & \mathsf{H}_{\Lambda}\\
	0 & \mathsf{K}_{\omega}\mathsf{H}_{\omega}^{1/2}\mathsf{B} & 0
	\end{pmatrix}\begin{pmatrix}\boldsymbol{\tau}\\
	\boldsymbol{\omega}\\
	\boldsymbol{\Lambda}
	\end{pmatrix}=\begin{pmatrix}\boldsymbol{\tau}_{\text{ext}}\\
	0\\
	\dot{\boldsymbol{\Lambda}}
	\end{pmatrix}}_{\text{strain-driven growth}}\quad\text{or}\quad\underbrace{\begin{pmatrix}\mathsf{B}^{T} & 0 & 0\\
	\mathsf{I}_{m} & -\mathsf{H}_{\omega}\mathsf{B} & \mathsf{H}_{\Lambda}\\
	\mathsf{K}_{\tau}\mathsf{H}_{\omega}^{-1/2} & 0 & 0
	\end{pmatrix}\begin{pmatrix}\boldsymbol{\tau}\\
	\boldsymbol{\omega}\\
	\boldsymbol{\Lambda}
	\end{pmatrix}=\begin{pmatrix}\boldsymbol{\tau}_{\text{ext}}\\
	0\\
	\dot{\boldsymbol{\Lambda}}
	\end{pmatrix}}_{\text{stress-driven growth}}.
\label{eq:linear-system}
\end{equation}
where the square diagonal real matrices with only positive entries $\mathsf{H}_{\omega},\, \mathsf{H}_{\Lambda}\in \mathbb{R}^{m\times m}$ result from linearising the constitutive law
\begin{equation}
(\mathsf{H}_{\omega})_{ii}=\frac{h_{i}'(l_{i}^{*}/L_{i}^{*})}{L_{i}^{*}}>0,\quad(\mathsf{H}_{\Lambda})_{ii}=\frac{h_{i}'(l_{i}^{*}/L_{i}^{*})l_{i}^{*}}{(L_{i}^{*})^{2}}>0
\label{eq:constitutive-coefficients-1d}
\end{equation}
The square diagonal real matrices $\mathsf{K}_{\omega},\mathsf{K}_{\tau},\,\in\mathbb{R}^{m\times m}$ are the growth constants which characterise how fast a cell grows in relation to the others,
\begin{equation}
(\mathsf{K}_{\omega})_{ii}=(\mathsf{H}_{\omega})_{ii}^{1/2}\left.\frac{\partial(\text{\ensuremath{\mathcal{G}}}_{\omega})_{i}}{\partial l_{i}}\right|_{\varepsilon=0},\qquad(\mathsf{K}_{\tau})_{ii}=(\mathsf{H}_{\omega})_{ii}^{-1/2}\left.\frac{\partial(\text{\ensuremath{\mathcal{G}}}_{\tau})_{i}}{\partial T_{i}}\right|_{\varepsilon=0}\,,\label{eq:growth-rates}
\end{equation}
where $(\mathcal{G}_{\omega})_{i}=(\mathcal{G}_{\omega})_{i}(l_{i})$ is a strain-driven growth law and $(\mathcal{G}_{\tau})_{i}=(\mathcal{G}_{\tau})_{i}(T_{i})$ is a stress-driven growth law. $\mathsf{I}_m$ is an $m \times m$ identity matrix. In both cases, $\boldsymbol{\tau}$ and $\boldsymbol{\omega}$ can be eliminated,
\begin{align}
\boldsymbol{\omega} & =(\mathsf{B}^{T}\mathsf{H}_{\omega}\mathsf{B})^{-1}(\mathsf{B}^{T}\mathsf{H}_{\Lambda}\boldsymbol{\Lambda}+\boldsymbol{\tau}_{\mathrm{ext}})\label{eq:omega-eliminated}\\
\boldsymbol{\tau} & =\left(\mathsf{H}_{\omega}\mathsf{B}(\mathsf{B}^{T}\mathsf{H}_{\omega}\mathsf{B})^{-1}\mathsf{B}^{T}-\mathsf{I}_{m}\right)\mathsf{H}_{\Lambda}\boldsymbol{\Lambda}+\mathsf{H}_{\omega}\mathsf{B}(\mathsf{B}^{T}\mathsf{H}_{\omega}\mathsf{B})^{-1}\boldsymbol{\tau}_{\mathrm{ext}}\label{eq:tau-eliminated}
\end{align}
leading to the linear system \eqref{eq:linear-system}. Note that
$\mathsf{B}^{T}\mathsf{H}_{\omega}\mathsf{B}$ is invertible, as explained in
Section \ref{sec:growth-dynamics-of-a-cell-network}. The growth dynamics of either growth	law in the neighborhood of the equilibrium state $\boldsymbol{w}^*$,
$\boldsymbol{T}^*$, $\boldsymbol{L}^*$ is linearly stable if the sign of the
largest non-zero eigenvalue of the respective Jacobian is negative. A
particularly helpful feature of the incidence matrix formulation of this
mechanical problem is the emergence of orthogonal projectors.
We define 
\begin{equation}
\mathsf{C}=\mathsf{H}_{\omega}^{1/2}\mathsf{B}\in\mathbb{R}^{m\times n}
\end{equation}
of full rank as well as
\begin{equation}
\mathsf{P}=\mathsf{C}(\mathsf{C}^{T}\mathsf{C})^{-1}\mathsf{C}^{T}\in\mathbb{R}^{m\times m}.
\label{eq:orthogonal-projector}
\end{equation}
We note that $\mathsf{P}$ is the orthogonal projector $\mathsf{P}$ onto the range of $\mathsf{C}$, satisfying $\mathsf{P}^{2}=\mathsf{P}^{T}=\mathsf{P}$, whereas $\mathsf{I}_{m}-\mathsf{P}$ is the orthogonal projector onto the null space of $\mathsf{C}^T$.

With wall displacements $\boldsymbol{\omega}$ and forces $\boldsymbol{\tau}$ eliminated according to \eqref{eq:omega-eliminated} and \eqref{eq:tau-eliminated}, the growth dynamics of both strain-driven and stress-driven laws in the neighborhood of the equilibrium state $\boldsymbol{\omega}^{*}$, $\boldsymbol{T}^{*}$, $\boldsymbol{L}^{*}$ can be written as
\begin{align}
&\dot{\boldsymbol{\Lambda}}=\mathsf{J}_{\omega}\boldsymbol{\Lambda},\quad\mathsf{J}_{\omega}=\mathsf{K}_{\omega}\mathsf{P}\mathsf{D}\label{eq:linear-dynamics1}\\
&\dot{\boldsymbol{\Lambda}}=\mathsf{J}_{\tau}\boldsymbol{\Lambda},\quad \mathsf{J}_{\tau}=-\mathsf{K}_{\tau}(\mathsf{I}_{m}-\mathsf{P})\mathsf{D}.
\label{eq:linear-dynamics}
\end{align}
where $\mathsf{J}_\omega$ and $\mathsf{J}_\tau$ are the Jacobians of the strain-driven and stress-driven cases, respectively. Here
\begin{equation}
\mathsf{D}=\mathsf{H}_{\omega}^{-1/2}\mathsf{H}_{\Lambda}
\end{equation}
is a positive diagonal matrix encoding the constitutive laws. 
There are three distinct contributions to the linearized dynamics: the diagonal matrices $\mathsf{K}_{\omega}$ and $\mathsf{K}_{\tau}$ depend on the growth law; the positive diagonal matrix $\mathsf{D}$ is obtained from  the constitutive law; and $\mathsf{P}$ encodes the network topology.
The special form of these Jacobian matrices implies that their eigenvalues are real, hence: \textit{The dynamics of both stress-driven and strain-driven growth laws cannot have oscillations.}

Next, we discuss under which conditions a growth law produces stable tissue-level dynamics (i.e. the largest non-zero eigenvalue of the respective Jacobian matrix is negative). We identify three important parameters: the number $k$ of positive  entries of either $\mathsf{K}_\omega$ or $\mathsf{K}_\tau$, the number of cells $m$, and the number of walls $n$ (not counting the leftmost wall).  Then, independently of the constitutive law or the topology of the network we have: \textit{Strain-driven growth is stable if all growth constants are negative ($k=0$). Stress-driven growth is stable if all growth constants are positive ($k=m$).} 

Similarly, we can identify cases where the dynamics is unstable: \textit{Strain-driven growth is unstable if $k+n>m$. Stress-driven growth is unstable if $k+n<m$. This is true for all constitutive laws and network topologies.} These two results thus generalize to an arbitrary network the link between growth constant and stability of an isolated cell.

\begin{figure}[t]
	\centering
	\includegraphics[width=0.8 \textwidth]{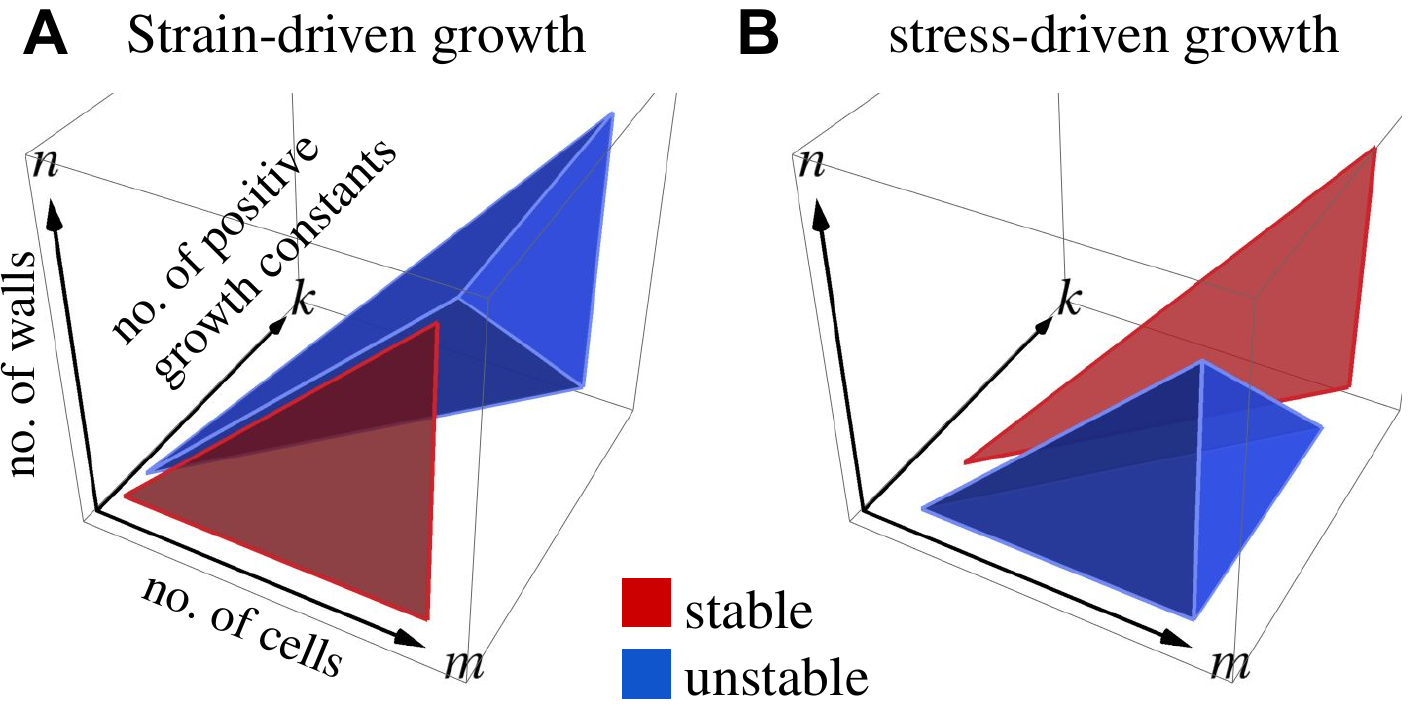}
	\caption{Parameter regions in which growth dynamics is guaranteed to be either stable (red) or unstable (blue) for strain-driven growth  (\textbf{A}) and stress-driven growth (\textbf{B}).  In between the stable and unstable parameter regions, stability depends on a particular choice of constitutive law and network topology.} 
	\label{fig2:pyramids-analytical}
\end{figure}

\section{\label{sec:proofs}Stability of growth dynamics for arbitrary network topologies, constitutive laws and growth laws}

We also have the  topological constraint $m \geq n$ (the network cannot have holes) and $0\leq k\leq m$ (there are between 0 and $m$ cells with positive growth constants). Therefore, in the  $(m, k, n)$ parameter space the sufficient conditions for stability and instability delineate two triangular pyramids, bounded by the plane $k+n=m$, where stability or instability will occur (indicated in red and blue Fig.~\ref{fig2:pyramids-analytical}).

We now proceed to provide proofs for the three propositions written in italics
in the main text (in the following, they are stated as Propositions \ref{prop:no-oscillations},
\ref{prop:stable-growth-dynamics} and \ref{prop:unstable-growth-dynamics}). They
relate to the non-oscillatory nature of the growth dynamics, and provide
necessary conditions for stable and unstable growth dynamics independently of
topology, constitutive laws and growth laws.

We consider an orthogonal projector $\mathcal{P}\in \mathbb{R}^{m\times m}$,
which is idempotent and symmetric: $\mathcal{P}^2=\mathcal{P}=\mathcal{P}^T$.
Let the real diagonal matrix $\mathcal{K}\in\text{diag}(\mathbb{R}^{m})$ have
$k$ positive entries. Further,
$\mathsf{D}\in\text{diag}\left(\mathbb{R}_{>0}^{m}\right)$ is a diagonal matrix
of positive entries (this follows from the requirement that constitutive laws
are monotonic, see (2)). Note that $\mathsf{D}$ is the same matrix that appears
in $(4)_2$ and $(5)_2$, whereas the form of the orthogonal projector $\mathcal{P}$ and
growth constants $\mathcal{K}$ vary depending on the growth law in consideration.
Let us also denote by $\Pi(\mathcal{K})$ the spectrum of $\mathcal{K}$
(i.e., the set of all eigenvalues of $\mathcal{K}$).

Firstly, our goal is to prove that the growth dynamics of both growth laws
cannot have oscillatory dynamics. To this end, we first prove a lemma about the
matrix $\mathcal{K}\mathcal{P}\mathsf{D}$ which mirrors the structure of both
growth laws.
\begin{lemma}
	\label{lemma:spectrum}
	The eigenvalues of the nonsymmetric matrix
	$\mathcal{J}=\mathcal{K}\mathcal{P}\mathsf{D}$ are real.
\end{lemma}
\textit{Proof.}
It follows from \cite[Thm. 1.3.22]{Horn2013MatrixAnalysis} that if
$\mathcal{P}_1$ and $\mathcal{P}_2$ are two square matrices,
then 
\begin{equation}
\Pi(\mathcal{P}_1\mathcal{P}_2)=\Pi(\mathcal{P}_2\mathcal{P}_1).
\end{equation}
Hence, since $\mathcal{K}$, $\mathcal{P}$ and $\mathsf{D}$ are square matrices,
and using the fact that $\mathcal{P}=\mathcal{P}^2=\mathcal{P}^T$, and $\mathsf{D}$ and
$\mathcal{K}$ commute,
we have that
\begin{equation}
\Pi(\mathcal{K}\mathcal{P}\mathsf{D})=\Pi(\mathsf{D} \mathcal{K} \mathcal{P}) = \Pi(\mathsf{D} \mathcal{K} \mathcal{P}^2) = \Pi(\mathsf{P}^T \mathsf{D} \mathcal{K} \mathcal{P}) = \Pi(\mathcal{P}^T \mathcal{K} \mathsf{D} \mathcal{P}).
\label{eq:spectrum}
\end{equation}
This shows that $\mathcal{K}\mathcal{P}\mathsf{D}$ and
$\mathcal{P}^{T}\mathcal{K}\mathsf{D}\mathcal{P}$ have the same eigenvalues
(note that $\mathcal{K}\mathsf{D}$ is a diagonal matrix). Since
$\mathcal{P}^{T}\mathcal{K}\mathsf{D}\mathcal{P}$ is symmetric, the eigenvalues
of the nonsymmetric matrix $\mathcal{K}\mathcal{P}\mathsf{D}$ are real.\hfill
$\square$ \\
\\
\begin{proposition}\label{prop:no-oscillations}
	The dynamics of both stress-driven and strain-driven growth laws cannot have oscillations.\end{proposition}

\textit{Proof.}
For the strain-driven growth law, the Jacobian is given by (4)$_2$, that is
$\mathsf{K}_\omega \mathsf{P} \mathsf{D}$, where $\mathsf{K}_\omega$ describes
the growth constants \eqref{eq:growth-rates}$_1$ and $\mathsf{P}$ is the orthogonal
projector as defined in \eqref{eq:orthogonal-projector}. Thus Lemma
\ref{lemma:spectrum} applies with $\mathcal{K}=\mathsf{K}_\omega$ and
$\mathcal{P}=\mathsf{P}$. For the stress-driven case, the Jacobian (5)$_2$ is
$-\mathsf{K}_\tau(\mathsf{I}_m-\mathsf{P})\mathsf{D}$ where $\mathsf{K}_\tau$
are the growth constants \eqref{eq:growth-rates}$_2$. Once again
Lemma~\ref{lemma:spectrum} applies with $\mathcal{K}=-\mathsf{K}_\tau$ and
$\mathcal{P}=\mathsf{I}_m -\mathsf{P}$. It follows for both growth laws that the
eigenvalues of the Jacobians are real, and no oscillations are possible. \hfill
$\square$ \\ \\

In what follows, we order the eigenvalues of an $m\times m$ matrix
$\mathcal{J}$ with only real eigenvalues in increasing order, i.e.,
$\sigma_{\text{min}}(\mathcal{J})=
\sigma_{1}(\mathcal{J})\leq\sigma_{2}(\mathcal{J})\leq\ldots\leq
\sigma_{m}(\mathcal{J})=\sigma_{\text{max}}(\mathcal{J})$.
We will need the following corollary to Ostrowski's theorem (see \cite[Cor. 4.5.11]{Horn2013MatrixAnalysis}).
\begin{corollary}
	\label{cor:Ostrowski}
	Let $\mathsf{Y}\in\mathbb{R}^{m\times m}$ be symmetric and let
	$\mathsf{X}\in\mathbb{R}^{m\times m}$.
	Then
	\begin{equation}
	\sigma_{j}(\mathsf{X}^{T}\mathsf{Y}\mathsf{X})=\theta_{j}\sigma_{j}(\mathsf{Y}),\qquad j=1,\ldots,m
	\end{equation}
	for some nonnegative real scalar $\theta_j$ such that
	$\sigma_{1}(\mathsf{X}^{T}\mathsf{X})\leq\theta_{j}\leq\sigma_{m}(\mathsf{X}^{T}\mathsf{X})$.
\end{corollary}

\begin{proposition}\label{prop:stable-growth-dynamics}Strain-driven growth is stable if all growth constants are negative ($k=0$). Stress-driven growth is stable if all growth constants are positive ($k=m$).\end{proposition}
\textit{Proof. }
Applying Corollary \ref{cor:Ostrowski} to $\mathsf{X}=\mathcal{P}$ and
$\mathsf{Y}=\mathcal{K}\mathsf{D}$, and using Lemma \ref{lemma:spectrum} leads
to
$\sigma_{j}(\mathcal{K}\mathcal{P}\mathsf{D})=
\theta_{j}\sigma_{j}(\mathcal{K}\mathsf{D})$
with
$\sigma_{1}(\mathcal{P}^{T}\mathcal{P})\leq\theta_{j}
\leq\sigma_{m}(\mathcal{P}^{T}\mathcal{P})$.
From the properties of $\mathcal{P}$, i.e.
$\mathcal{P}^{2}=\mathcal{P}^{T}=\mathcal{P}$, it follows that $\mathcal{P}$ has
eigenvalues 0 or 1, thus $0\leq\theta_{j}\leq1$ for all $j=1,\ldots,m$. In the
case that $\mathcal{K}$ has only non-positive entries, so does
$\mathcal{K}\mathsf{D}$ and we denote $\mathcal{K}$ with only non-positive
entries as $\mathcal{K}_{\leq0}$. Then
\begin{equation}
0\geq\sigma_{\text{max}}(\mathcal{K}_{\leq0}\mathcal{P}\mathsf{D})\geq\sigma_{\text{max}}(\mathcal{K}_{\leq0}\mathsf{D}).
\label{eq:Ostrowski-inequality}
\end{equation}
For the strain-driven growth law ($\mathcal{K}=\mathsf{K}_\omega$, $\mathcal{P}=\mathsf{P}$), it follows from  Eq.~\eqref{eq:Ostrowski-inequality} that the strain-driven growth with non-positive growth constants $\mathsf{K}_{\leq0}$ leads to a stable equilibrium irrespective of the network topology encoded in $\mathsf{P}.$

For the stress-driven growth law ($\mathcal{K}=-\mathsf{K}_\tau$, $\mathcal{P}=\mathsf{I}_m-\mathsf{P}$), we note that $\sigma_{\mathrm{max}}(-\mathsf{K}_{\geq0}\mathsf{D})=-\sigma_{\mathrm{min}}(\mathsf{K}_{\geq0}\mathsf{D})$. Eq.~\eqref{eq:Ostrowski-inequality} then becomes
\begin{equation}
0\geq\sigma_{\mathrm{max}}(\mathsf{J}_{\tau})
\geq-\sigma_{\mathrm{min}}\big((\mathsf{K}_{\tau})_{\geq0}\mathsf{D}\big)
\label{eq:Ostrowski-inequality-stress-driven}
\end{equation}
and it follows from  Eq.~\eqref{eq:Ostrowski-inequality-stress-driven} that the
stress-driven growth with non-negative growth constants $\mathsf{K}_{\geq0}$ leads
to a stable equilibrium irrespective of the network topology encoded in
$\mathsf{P}.$  \hfill $\square$\\ \\
The following proposition prepares our claims about unstable growth dynamics,
which are stated in the subsequent Proposition
\ref{prop:unstable-growth-dynamics}. \\

\begin{proposition} \label{prop:min-max}
	Let $\mathcal{P}\in\mathbb{R}^{m\times m}$ be an orthogonal projector of rank
	$n\le m$, $\mathcal{K}\in\mathbb{R}^{m\times m}$ be diagonal with $k$
	positive eigenvalues, and $\mathsf{D}\in\mathbb{R}^{m\times m}$ be diagonal
	with positive diagonal entries.
	If $k>m-n$ then the largest eigenvalue of $\mathcal{K}\mathcal{P}\mathsf{D}$ is
	positive.
\end{proposition}

\textit{Proof.}
Recall from \eqref{eq:spectrum} that $\mathcal{K}\mathcal{P}\mathsf{D}$ and
$\mathcal{P}^{T}\mathcal{K}\mathsf{D}\mathcal{P}$ have the same eigenvalues.
Then
$$
\sigma_{\max}(\mathcal{K}\mathcal{P}\mathsf{D})
=\sigma_{\max}(\mathcal{P}^{T}\mathcal{K}\mathsf{D}\mathcal{P})
=\max_{\left\Vert x\right\Vert _{2}=1}x^{T}\mathcal{P}^{T}\mathcal{K}\mathsf{D}\mathcal{P}x
=\max_{{y\in\text{range}(\mathcal{P})\atop \left\Vert y\right\Vert _{2}=1}}y^{T}\mathcal{K}\mathsf{D}y\,.
$$
Since $\dim(\text{range}(\mathcal{P}))=\text{rank}(\mathcal{P})=n$,
$$
\max_{{y\in\text{range}(\mathcal{P})\atop \left\Vert y\right\Vert _{2}=1}}y^{T}\mathcal{K}\mathsf{D}y\geq\min_{\dim V=n}\max_{{y\in V\atop \left\Vert y\right\Vert _{2}=1}}y^{T}\mathcal{K}\mathsf{D}y=\sigma_{n}(\mathcal{K}\mathsf{D}),
$$
where the last equality follows from the Courant-Fischer min-max theorem for
symmetric matrices~\cite[Thm.~2.4.6]{Horn2013MatrixAnalysis}. Since $n>m-k$, we
have that $n\geq m-k+1$ so that
$\sigma_{n}(\mathcal{K}\mathsf{D})\geq\sigma_{m-k+1}(\mathcal{K}\mathsf{D})$.
But $\mathcal{K}$, as well as $\mathcal{K}\mathsf{D}$ have $k$ positive
eigenvalues so $\sigma_{m-k+1}(\mathcal{K}\mathsf{D})>0$. Hence we conclude that
$\sigma_{\max}(\mathcal{K}\mathcal{P}\mathsf{D})\geq\sigma_{n}(\mathcal{K}\mathsf{D})>0$.
\hfill $\square$

\begin{proposition}\label{prop:unstable-growth-dynamics}Strain-driven growth is unstable if $k+n>m$. Stress-driven growth is unstable if $k+n<m$.\end{proposition}

\textit{Proof.}
The strain-driven case follows from applying Proposition \ref{prop:min-max} with $\mathcal{K}=\mathsf{K}_{\omega}$ and $\mathcal{P}=\mathsf{P}$. For the stress-driven case, Proposition \ref{prop:min-max} can be applied to the Jacobian $-\mathsf{K}_{\tau}(\mathsf{I}_{m}-\mathsf{P})\mathsf{D}$  with $\mathcal{K}=-\mathsf{K}_{\tau}$
and $\mathcal{P}=\mathsf{I}_{m}-\mathsf{P}$. Indeed, $\mathsf{I}_{m}-\mathsf{P}$
is the orthogonal projector onto the orthogonal complement of
$\text{range}(\mathsf{C})$, which
is the null space of $\mathsf{C}^T$.
So $\text{rank}(\mathsf{I}_{m}-\mathsf{P})=
\dim\big(\text{range}(\mathsf{I}_{m}-\mathsf{P})\big)=
\dim\big((\text{null}(\mathsf{C}^T)\big)=m-n$.
Since $-\mathsf{K}_\tau$ has $m-k$ positive eigenvalues, it follows from Proposition
\ref{prop:min-max} that if $k<m-n$ then the largest eigenvalue of
$-\mathsf{K}_{\tau}(\mathsf{I}_{m}-\mathsf{P})\mathsf{D}$ is positive. \hfill
$\square$




\section{Some cells try to grow, others shrink: When stability depends on specifics }

We now turn our attention to the intermediate region between the stable (red) and unstable (blue)  parameter regions in Fig.~\ref{fig2:pyramids-analytical}. In this case, stability depends on the  specific network topology, constitutive law, and growth law. Here, we choose a Hencky-type constitutive law \cite{mihai2017characterize} $h_i=\mu_i \log(\alpha_i)$. Our goal is to evaluate numerically $\mathsf{J}_\omega$, $\mathsf{J}_\tau$  to extract their eigenvalues for a given network topology $\mathsf{B}$.

To construct the Jacobians, two challenges must be met: First, a network topology $\mathsf{B}$ must be generated. Secondly, using $\mathsf{B}$, an equilibrium state given by $\boldsymbol{w}^*$, $\boldsymbol{T}^*$ and $\boldsymbol{L}^*$ can be generated. These two steps allow us to compute all the ingredients for the Jacobians $\mathsf{J}_\omega$ and $\mathsf{J}_\tau$. We now provide some details on the steps.


\subsection{\label{secSI:geometry-generation}Random generation of reference network geometries}

\begin{figure*}
	\includegraphics[width=1 \textwidth]{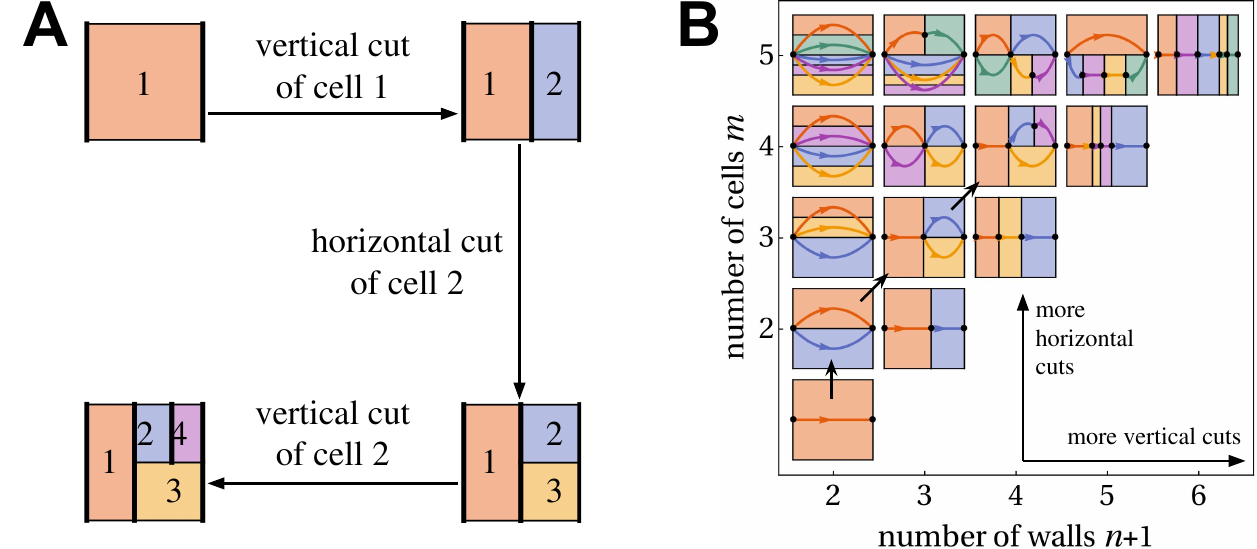}
	\centering
	\caption{Illustration of the cell division algorithm for creating random network topologies. \textbf{A} A network is created by making sequential cuts from a unit cell. Before every cut, a cell is picked at random. The cutting operation is also picked at random, either a horizontal or a vertical cut. Thus a new rectangular cell emerges. This process is repeated until the final network is generated. \textbf{B} Randomly generated topologies, ordered by number of walls and number of cells. Moving to the right corresponds to adding a vertical cut, moving upwards adds a horizontal cut. Graph representations are shown as overlays.}
	\label{fig:many-topologies}
\end{figure*}

The stress-free reference geometry (Fig. \ref{fig1:morphoelastic-bar}A) is created by taking a rectangular unit cell as a starting point, and then dividing it in a sequence of horizontal and vertical cuts. Fig.~\ref{fig:many-topologies}A demonstrates the process. Starting with a single cell (a square, cell 1) we decide randomly whether to make a horizontal or vertical cut. From the resulting two cells (red and blue), we pick a cell at random (in this case cell 2) and randomly decide one of two cut operations (in this case, a horizontal cut). The process is repeated until the final network has been reached. Random numbers are drawn from a uniform distribution using the function $\texttt{RandomInteger}$ in Wolfram Mathematica$^\text{\textregistered}$ 11.2.

Fig.~\ref{fig:many-topologies}B shows a number of randomly generated topologies, sorted by their number of cells and walls. Each topology has an overlay of the graph representation. Moving one unit upward on this grid jumps to a network with one more horizontal cut (i.e. series connection). Moving one unit to the right adds a vertical cut (i.e. parallel connection). The path along which cuts from Fig.~\ref{fig:many-topologies}A have been made is shown with black arrows.
\chg{The algorithm presented here of making sequential vertical and horizontal cuts places a restriction on the tiling geometries, and does not cover all possible tilings. This algorithm is used for the numerical results in the following section and in Fig.~ \ref{fig3:pyramids-numerical}. However, we note that the results of Section \ref{sec:proofs} and Fig.~\ref{fig2:pyramids-analytical} are not affected by this restriction of tilings, as they do not depend on the explicit form of the incidence matrix. }

\subsection{\label{secSI:numerical-stability}Numerical linear stability analysis for a non-linear constitutive law}

\begin{figure}[t]
	\centering
	\includegraphics[width=0.8 \textwidth]{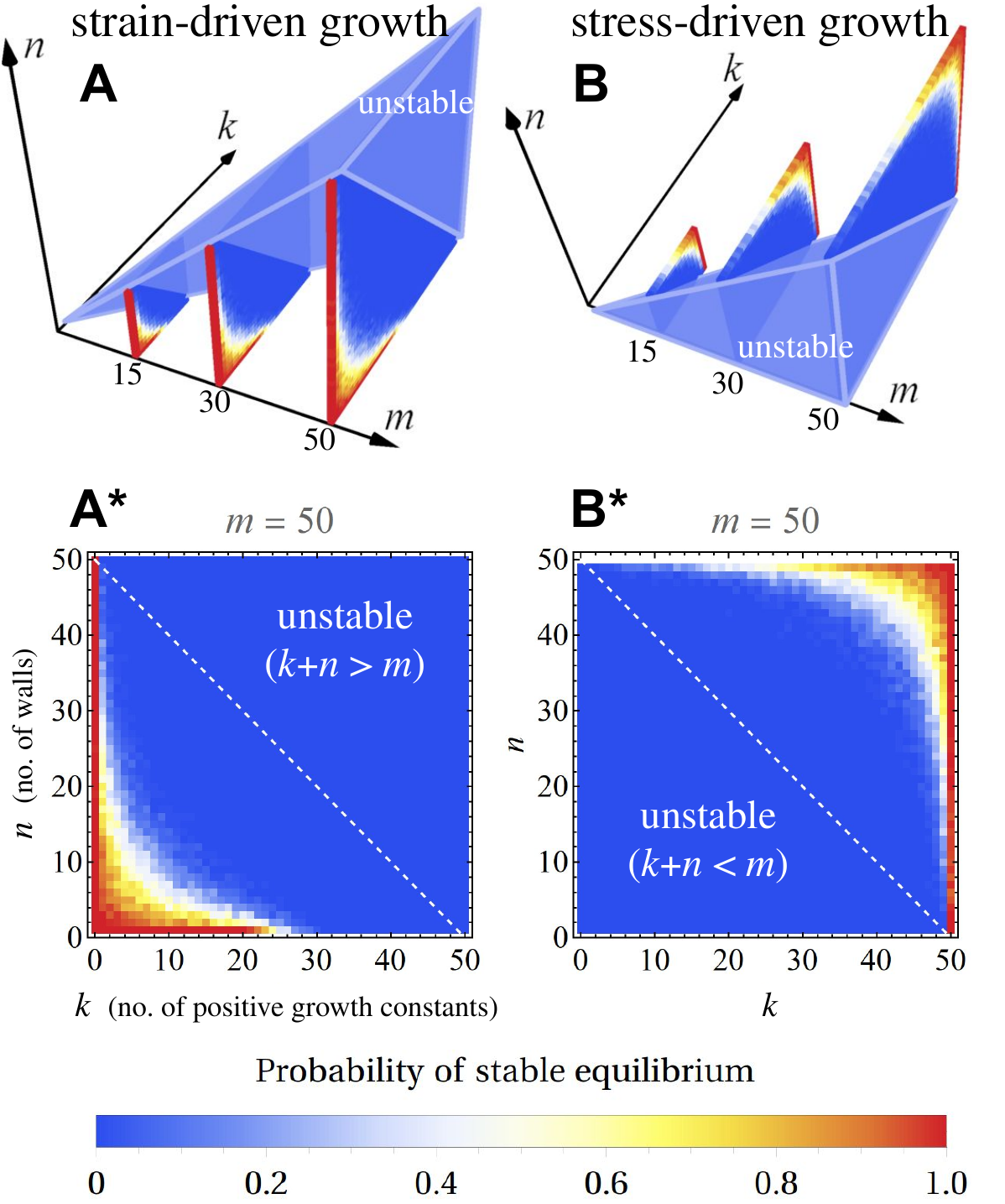}
	\caption{Probability of  stable equilibria in randomized networks. The color of each pixel is given by the probability of stable growth calculated from 100 simulated Jacobian matrices, obtained from randomly generated continuous parameters (e.g. lengths of reference network, growth constants) and a random network topology. \textbf{A} and \textbf{B} show probabilities of a stable equilibrium for stress-driven and strain-driven growth (the largest non-zero eigenvalues of the respective Jacobian matrices are negative), respectively, for up to 50 cells. The pyramidal region of unstable growth from Fig.~\ref{fig2:pyramids-analytical} is overlaid. \textbf{A*} and \textbf{B*} show detailed views for $m=50$ cells. In the strain-driven case (\textbf{A*}), parallel connections are more likely to produce a stable configuration. In the stress-driven case (\textbf{B*}), series connections are more likely to produce a stable configuration.}
	\label{fig3:pyramids-numerical}
\end{figure}

Here we demonstrate how for a specific growth law and constitutive law, a randomly generated numerical Jacobian for a stress-driven or strain-driven growth law is obtained. This random generation of Jacobians underlies Fig. \ref{fig3:pyramids-numerical}.

We focus on a Hencky type constitutive law $h_{i}(\alpha_{i})=\mu_{i}\log\alpha_{i}$. For a strain-driven growth law, we choose 
\begin{equation}
\boldsymbol{\mathcal{G}}_{\omega}=\mathsf{K}_{\omega}(\boldsymbol{l}-\boldsymbol{l}^{*})=\mathsf{K}_{\omega}\mathsf{B}(\boldsymbol{w}-\boldsymbol{w}^{*}).
\end{equation}
The stress-driven growth law is
\begin{equation}
\boldsymbol{\mathcal{G}}_{\tau}=\mathsf{K}_{\tau}(\boldsymbol{T}-\boldsymbol{T}^{*}).
\end{equation}
The first key challenge is to obtain a randomly generated equilibrium state given by $\boldsymbol{w}^{*}$, $\boldsymbol{T}^{*}$ and $\boldsymbol{L}^{*}$ for the non-linear Hencky constitutive law. This requires a numerical solution of \eqref{eq1:KCL} and \eqref{eq2:constitutive-single-cell}. By definition of an equilibrium state $\dot{\boldsymbol{L}}^{*}=0$, the growth dynamics trivially satisfies $\boldsymbol{\mathcal{G}}_{\omega}(\boldsymbol{l}^*)=\boldsymbol{\mathcal{G}}_{\tau}(\boldsymbol{T}^*)=0$. The remaining equations are the force balance and constitutive laws,
\begin{equation}
\mathsf{B}^{T}\boldsymbol{T}^{*}=\boldsymbol{T}_{\mathrm{ext}},\qquad\boldsymbol{T}^{*}=\text{diag}(\boldsymbol{\mu})\left[\log\left(\mathsf{B}\boldsymbol{w}^{*}\right)-\log\left(\boldsymbol{L}^{*}\right)\right].\label{eq:balance-laws-Hencky}
\end{equation}
To solve (\ref{eq:balance-laws-Hencky}), we prescribe a growth profile $\boldsymbol{L}^{*}\in\mathbb{R}^{m}$ which we obtain from a uniform random distribution. The reference lengths $\boldsymbol{L}^{*}$ correspond to the individual cells' preferred sizes. We also obtain a reduced incidence matrix from a random graph, obtained through a procedure described in Section \ref{secSI:numerical-stability}. Thus $\boldsymbol{w}^{*}$ and $\boldsymbol{T}^{*}$ can be obtained through non-linear root finding. Once they are known, we calculate $\mathsf{H}_{\omega}$ and $\mathsf{H}_{\Lambda}$ according to Eq.~\eqref{eq:constitutive-coefficients-1d}:
\begin{equation}
\mathsf{H}_{\omega}=\text{diag}(\boldsymbol{\mu})\,\text{diag}^{-1}(\mathsf{B}\boldsymbol{w}^{*}),\qquad\mathsf{H}_{\Lambda}=\text{diag}(\boldsymbol{\mu})\,\text{diag}^{-1}(\boldsymbol{L}^{*}).
\end{equation}
Then the matrices $\mathsf{C}$ and $\mathsf{D}$ can be obtained as described in the main text. We randomly generate the growth constants $\mathsf{K}_{\omega}$, $\mathsf{K}_{\tau}$ drawing from a uniform distribution, ensuring with an extra constraint that $0\leq k\leq m$ of its of their entries are positive. Then $\mathsf{P}$, $\mathsf{J}_{\omega}$ and $\mathsf{J}_{\tau}$ can be obtained by \eqref{eq:orthogonal-projector}, \eqref{eq:linear-dynamics1} and \eqref{eq:linear-dynamics}. In summary, we have described how an equilibrium state compatible with the constraints of a Hencky type constitutive law can be constructed. This allows us to compute all the ingredients needed for the Jacobians $\mathsf{J}_{\omega}$ and $\mathsf{J}_{\tau}$, which tells us the stability of growth dynamics in the neighborhood of a randomly generated equilibrium state $\boldsymbol{w}^{*}$, $\boldsymbol{T}^{*}$, $\boldsymbol{L}^{*}$.

\section{How likely is it that an equilibrium state in a randomly generated network leads to stable dynamics?}

\begin{figure*}
	\includegraphics[width=0.8\textwidth]{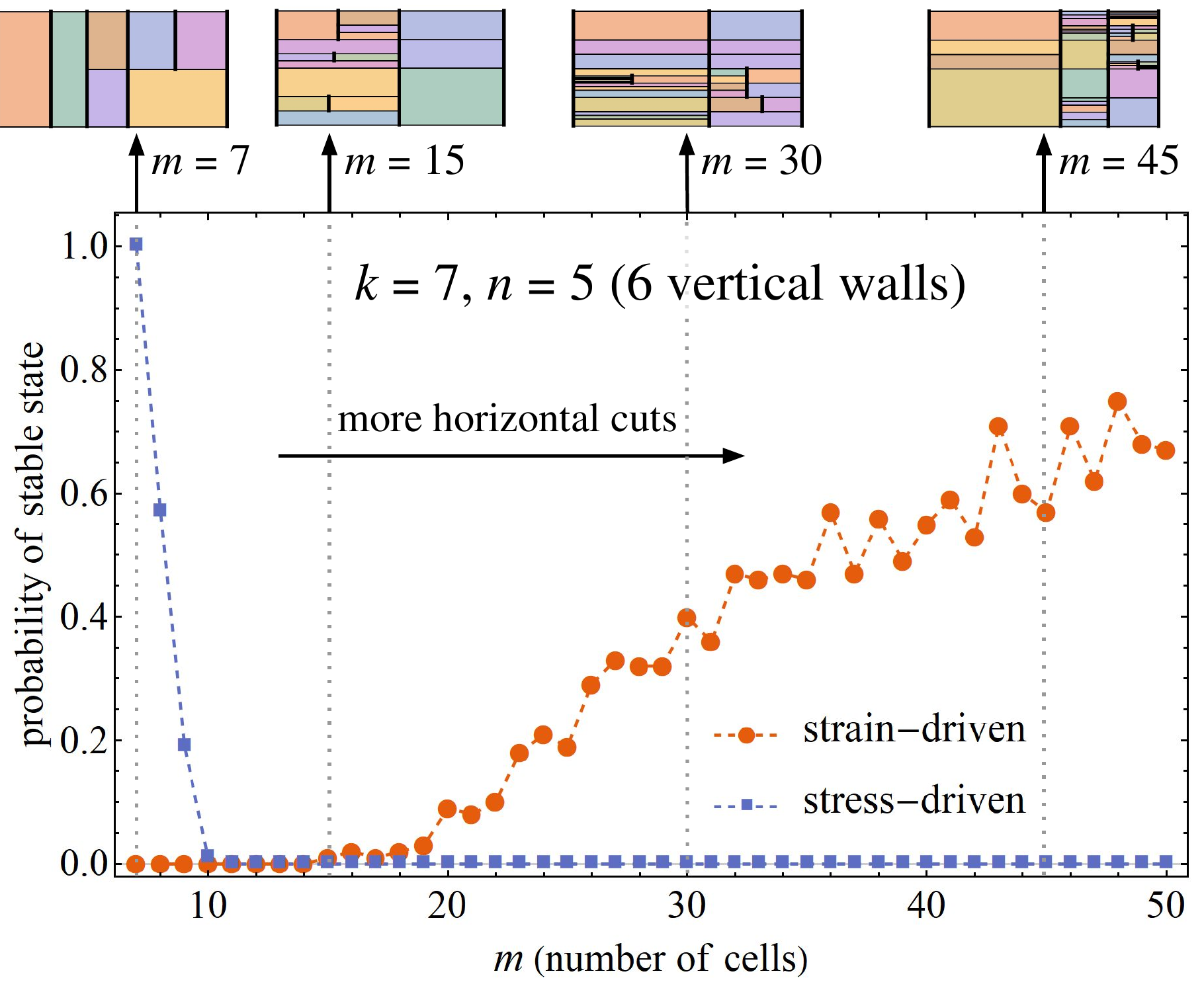}
	\centering
	\caption{One-dimensional slice through the data of Fig.~\ref{fig3:pyramids-numerical}),  showing how probabilities of stable growth dynamics change as the number of horizontal cuts, i.e. parallel connections, increases.}
	\label{fig:more-horizontal-cuts}
\end{figure*}

To quantify how likely it is that an equilibrium state for a given growth law is stable, we use Jacobian matrices constructed from randomly generated parameters as described above. We define the probability of a stable equilibrium as the number of randomly generated Jacobian matrices of which the largest non-zero eigenvalue is negative, divided by the total number of randomly generated Jacobian matrices. Fig.~\ref{fig3:pyramids-numerical}A and B show the probabilities of stable strain-driven and stress-driven growth dynamics, respectively, for up to 50 cells. For reference, the unstable regions from Fig.~\ref{fig2:pyramids-analytical} are superimposed. The color of each pixel corresponds to the probability of stable growth calculated from 100 randomly generated matrices. Fig.~\ref{fig3:pyramids-numerical}A* and B* show the fixed $m=50$ case in detail. The largest regions of high probability for stable strain-driven growth occur for small values of $n$ and $k$ compared to $m$ (bottom left corner of A*), implying that parallel connections are more likely to produce a stable configuration. Conversely, for stress-driven growth (top right corner of B*), series connections are more likely to produce a stable configuration. Fig.~\ref{fig:more-horizontal-cuts} confirms this via a one-dimensional slice of the data of  Fig.~\ref{fig3:pyramids-numerical}, showing how probabilities of stable growth dynamics change as the number of horizontal cuts, i.e. parallel connections, increases.

\begin{figure*}
	\hfill{}\includegraphics[width=1 \textwidth]{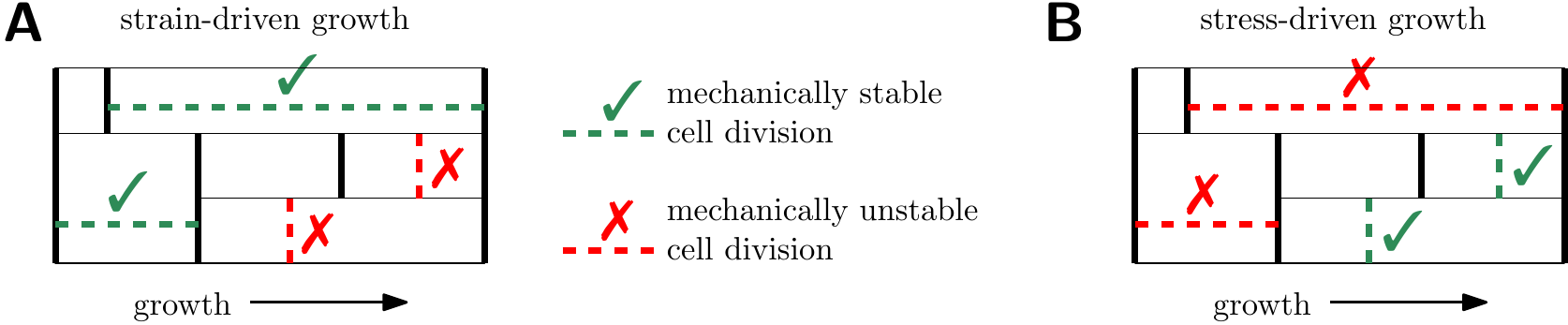}\hfill{}
	\caption{Cell division planes consistent with stable growth dynamics. Dashed lines indicate hypothetical division planes. Division along the green lines increase probability of stable growth dynamics according to Fig.~\ref{fig3:pyramids-numerical}, division along red lines decreases it. \textbf{A} For stain-driven growth, cell divisions planes parallel to the growth direction are favourable from a mechanical stability point of view when compared to perpendicular division planes. \textbf{B} The reverse is true in the stress-driven case: Division planes perpendicular to the growth direction are favourable.}
	\label{fig4:cell-division}
\end{figure*}

While we are not  tracking possible topological modifications such as cell division or neighbor exchanges, the results of Fig.~\ref{fig3:pyramids-numerical} allows us to compare the probability of stability before and after cell division. Assuming a growth dynamics model, we can make inferences about whether cell division planes are likely to be parallel or perpendicular to the direction of growth. Indeed, Fig.~\ref{fig3:pyramids-numerical} implies that in order for cell division to remain stable, planes parallel to the direction of growth are  associated with strain-driven growth, whereas cell division planes perpendicular to the direction of growth are  associated with stable stress-driven growth. 
The relationship between cell division planes and mechanical stability is summarised in Fig.~\ref{fig4:cell-division}.

\section{Discussion}

Our analysis suggests the following mechanism in uniaxial tissue growth dynamics: stress-driven growth laws lead to series connections at the level of the whole network while strain-driven growth leads to parallel connections. Our approach relates the topological structure of the system to its growth dynamics.  

\chg{We have developed our model under the geometric restriction that cells grow and deform in only one dimension. While this is a clear limitation of the model, the advantage of this simplification is an analytical description that allows us to make concrete statements, as for instance presented in Figs. \ref{fig2:pyramids-analytical} and \ref{fig3:pyramids-numerical}, that connect network topology and the local rules of growth dynamics at the cell level to the global dynamic response. Moreover, there exist biological systems that are reasonably well-approximated by such a system, for instance the structure of skeletal muscle. In muscles, sarcomeres have been modeled as one-dimensional active elements connected in a complex layered 1D network structure that forms myofibrils, muscle fibres, and further layers that build up the muscle tissue \cite{caruel2018physics}. Furthermore, in heart tissues, the addition of sarcomeres in series and parallel has been associated with different loading conditions, see \cite{goktepe2010athlete}, suggesting that there are microscopic mechanisms that locally change the network topology in response to different end conditions.} 

Another interesting application of these ideas can be found in the growth dynamics of plant stems and roots. For instance, the \textit{Arabidopsis thaliana} plant root is mostly a uni-directionally growing network where cells are primarily connected in series with respect to the growth direction. The growth dynamics is well captured by a standard physiological model by Lockhart, in which uni-directional plant cell growth is described as proportional to a difference between forces due to the cell's turgor pressure and a biologically encoded yield pressure, with a positive proportionality constant \cite{jensen2015multiscale}. Lockhart's law is  a stress-driven growth law with a positive growth constant. This model description, when paired with the observation that the \textit{Arabidopsis thaliana} is mostly comprised of series connections, is consistent with the criteria of stable stress-driven growth dynamics, and thus supports the hypothesis that tissues maximize the probability of stable growth dynamics. 

\chg{Aside from dimensionality, our model has several limitations that need to be addressed in a more comprehensive theory of the dynamics of growing biological tissue. Friction and/or adhesion between cells sharing a horizontal wall would be a useful addition, as cortical tension and cell adhesion have been recognised as important factors in epithelial monolayers \cite{farhadifar2007influence}. Also, cell wall anisotropy \cite{jensen2015multiscale} and intracellular fluxes \cite{cheddadi2019coupling} are of key importance in plants. A popular class of models for both types of tissue are vertex models \cite[p.~56]{goriely17}. 
	
	On the other hand, incorporating some of the components described above would add significant complexity to our model system, whereas the relative simplicity and analytical tractability may be helpful to study the qualitative impact of growth and mechanics in complex biological  problems, such as size regulation and growth termination, i.e. how the overall domain size of an organ or tissue (e.g. the wing disc of \textit{Drosophila melanogaster}) is communicated to the individual cells  to stop growth and division. Various cell-based \cite{shraiman2005mechanical,hufnagel2007mechanism} and continuum \cite{aegerter2007model,ambrosi2015active} models, incorporating a combination of mechanical feedback and morphogen diffusion, postulate different local growth laws and illustrate the complexity of finding a minimal mechanism that incorporates size regulation and growth termination. We believe that our simplified one-dimensional model, to which discrete modeling of advection and diffusion can be coupled, is  a first step for theoretical studies of size regulation, as finite equilibrium states of growth dynamics can be interpreted as states of growth termination.  }

\chg{In the light of recent work work by Jensen et al. \cite{jensen2019force}, an extension of our 1D model to 2D vertex model geometries (exploiting their algebraic structure with incidence matrices and discrete calculus) may be promising. Further extensions of our model to better approximate biological tissue (for instance for the study of growth termination in epithelia)  would include} advection, diffusion and dilution of chemical patterns \cite{aguilar2018critical,erlich2018physicalDeterminants,grady2010discrete} and would take account of the (near-) incompressibility of cells \cite{harris2012characterizing}, as well as dynamical topological changes caused by cell division \cite{xu2015changes}, which requires an accompanying topological growth law.

\vskip6pt

\enlargethispage{20pt}

\noindent\textbf{Ethics statement.} This work did not involve any active collection of human data.\\
\textbf{Data accessibility statement.} All data needed to evaluate the results and conclusions are present in the paper. The associated datasets and codes can be accessed via the Figshare repository (\url{https://doi.org/10.6084/m9.figshare.10315775}). \\
\textbf{Competing interests statement.} We have no competing interests.\\
\textbf{Authors' contributions.} AG, DEM, and AE devised the study, AE conducted the analysis, GWJ and AE devised the network description, FT and AE formulated the algebraic theorems and proofs, all authors contributed to the writing of the paper. \\
\textbf{Funding.} This work was supported by the Engineering and Physical Sciences Research Council grant EP/R020205/1 to Alain Goriely.
%
\appendix{

	\section{\label{sec:squaring-the-square}Cell diagrams and electrical analogy in the mathematical puzzle of Squaring the Square}
	
	A similar concept to cell diagrams was devised in the 1940s by four undergraduate students at Trinity College Cambridge, who were trying to find perfect squared squares. The idea is to dissect a square into smaller squares which do not repeat in size. Prior to the work of R.L. Brooks, C.A.B. Smith, A.H. Stone and W.T. Tutte which was published in the classic paper \emph{The dissection of rectangles into squares} \cite{Brooks1940}, a number of squared rectangles and imperfect squares (requiring repeated triangles) had already been known. The four students searched for perfect squared squares mostly empirically, but developed an extensive theory of squared rectangles which combined the theory of planar graphs and of electrical networks, and to which which cell diagrams and cell graphs have a lot of similarity. Their networks, which they called polar nets or p-nets, but which were later dubbed Smith diagrams after C.A.B. Smith, were simply connected polar directed acyclic graphs same as the cell diagrams presented in this work. The students took advantage of Kirchhoff's laws purely as a tool to balance lengths, interpreting the sides of squares as currents, and potential differences as lengths (with the assumption of unit resistance, and disregarding physical units). Smith diagrams were a semi-empirical tool to organise the imperfect squared squares found in the process towards discovering perfect squared squares, as described later in an essay by W.T. Tutte \cite{Tutte1961}. Later an exhaustive computer search revealed that the smallest squared square contains 21 squares, and this is the current logo of the Trinity Mathematical Society\footnote{Trinity Mathematical Society logo and its history: \url{https://tms.soc.srcf.net/about-the-tms/the-squared-square/\#2}}. Figure \ref{fig:squared-squares} shows the first published squared rectangle known as Lady Isabel's Casket, as well as the original illustration of p-nets in \cite{Brooks1940}. For a detailed history of squared squares and an overview over the use of p-nets, see \cite{Anderson2013}. For a concise popular overview see \cite{Stewart1995}. 
	
	\begin{figure} \includegraphics[width=1\textwidth]{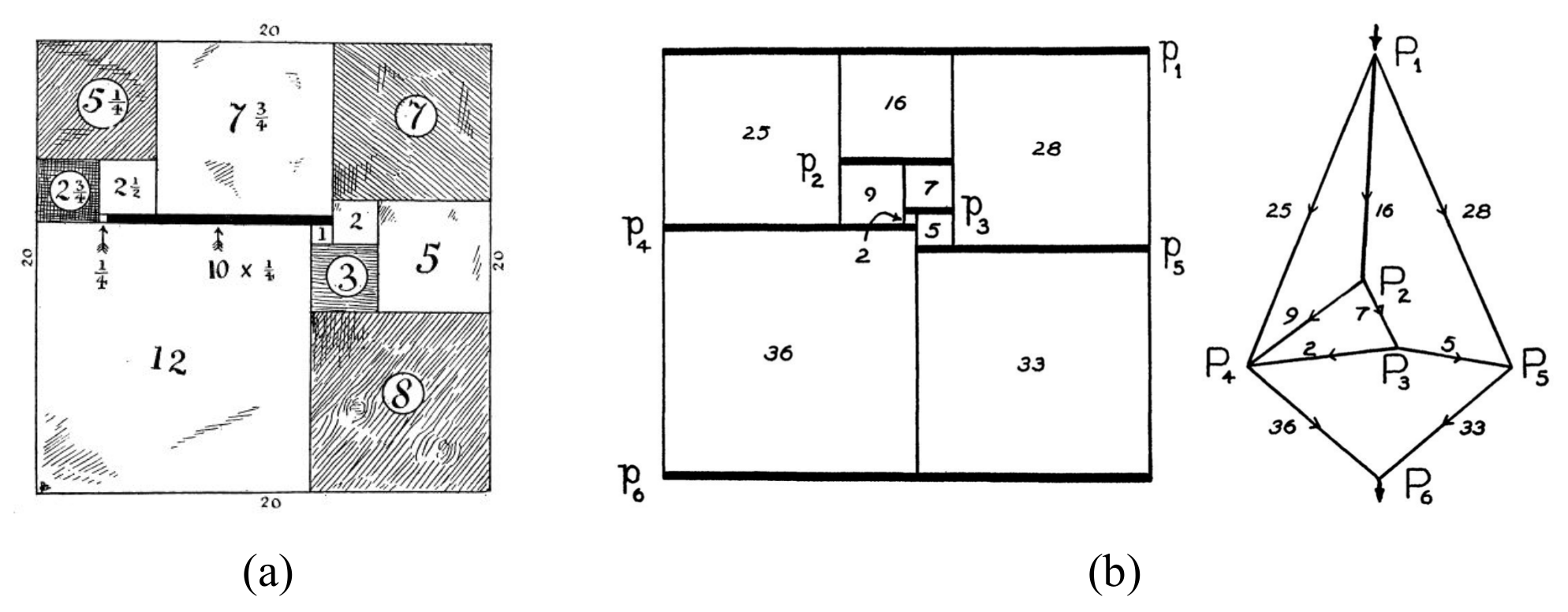}
		\centering
		\caption{\label{fig:squared-squares}(a) The first published reference dealing with the dissection of a square into smaller different sized squares was a recreational puzzle known as \emph{Lady Isabel's Casket}. Five years after its first publication by H.E. Dudeney, it was republished in 1907 in the collection \emph{The Canterbury Puzzles }which is still available in print today. (b) A squared rectangle and the associated p-net, as presented in \cite{Brooks1940}. The squared rectangle can be represented as an electrical circuit. Each horizontal line corresponds to a node. Currents flow into the top of the node and out of the bottom of one. The amount of current through each wire equals the size of the square represented. } 
	\end{figure}	
	
}

\newpage
\bibliographystyle{RS}

\end{document}